\documentclass[11pt]{article}
\usepackage{ams math,amssymb,graphicx,url}
\usepackage{color}

\newcommand\ignore[1]{\textcolor{red}{cut text:}}
\def\cut{\Lambda}

\renewcommand\vec{\boldsymbol}
\renewcommand\varphi{\phi}
\renewcommand\hat{\widetilde}

\begin{document}

\newpage 

\author{R. C. ~Brower$^a$, G. T. Fleming$^b$, H.~Neuberger$^c$
\footnote{Weston Visiting Scientist at The Weizmann Institute of Science.}\\ [3mm]
  {\normalsize\it \llap{$^a$} Department of Physics, Boston University}\\
  {\normalsize\it  Boston, Massachusetts 02215, USA}\\
  {\normalsize\tt brower@bu.edu}\\ [4mm]
  {\normalsize\it \llap{$^b$} Department of Physics, Sloane Laboratory, Yale University}\\
  {\normalsize\it   New Haven, Connecticut 06520, USA}\\
  {\normalsize\tt George.Fleming@yale.edu}\\ [4mm]
  {\normalsize\it \llap{$^c$}Department of Physics and Astronomy, Rutgers University}\\
  {\normalsize\it Piscataway, NJ 08855, U.S.A} \\
  {\normalsize\tt neuberg@physics.rutgers.edu}\\ [4mm]
}

\title{Lattice Radial Quantization: 3D Ising}

\date{December 26, 2012}

\maketitle 

\vskip 0.5cm

\begin{abstract} 
\noindent 
Lattice radial quantization is introduced as a nonperturbative method intended to numerically solve
Euclidean conformal field theories that can be realized as fixed points of known 
Lagrangians. As an example, we employ a lattice shaped as a cylinder with a
2D Icosahedral cross-section to discretize dilatations in the 3D Ising model.
Using this method, we obtain the preliminary estimate $\eta=0.034(10)$.

\end{abstract}

\bigskip
\newpage

\section{Introduction}

Conformal field theories are theoretically interesting in their own right,  have applications in
Condensed Matter and Statistical Physics and are relevant to Particle Physics in general. 
They may play a central role in the yet unknown correct description of 
Particle Physics beyond the Standard Model. Traditional numerical methods work well when
the correlation length can be kept at a few lattice spacings but are ill suited for 
conformal field theories. Here we propose an untraditional numerical formulation, based 
on radial quantization.

A classical local Euclidean action on $\mathbb{R}^d$  with no dimensional parameters can be rewritten 
in polar coordinates with radius $r$ as a field theory on $\mathbb{S}^{d-1}\times \mathbb{R}$ with invariance under 
shifts along the factor $\mathbb{R}$ parametrized by $\log(r)$~\cite{Fubini:1972mf}. 
The shift symmetry corresponds to
dilatations about the chosen origin in $\mathbb{R}^d$. 
The corresponding generator is the dilatation operator and its eigenvalues
are operator dimensions. 
The critical new feature of our proposal is to introduce a uniform lattice discretization of $\log(r)$; 
this provides a reasonable match between numerical resources and the relevant degrees of freedom.
A lattice of linear size $L$ in  $\log(r)$ in our proposal replaces a lattice of linear 
size $e^L$ when $\mathbb{R}^d$ is regularized directly.
Using the 3D Ising model, we carry out a modest numerical test of lattice radial
quantization of the Wilson-Fisher fixed point field theory.

\section{Preserving dilatation invariance}

It may sound surprising that one can regularize a field theory that is classically conformally invariant while 
preserving dilatations because this seems to say that an asymptotically free theory, such as QCD, does not 
necessarily generate a mass.
We first clear up this point by  studying the large N limit of the $O(N)$ sigma model in 2D, and in the course of 
doing so we learn how to identify a theory that does stay
conformal after quantization.

\subsection{An example with asymptotic freedom}

Consider the quantization of the two
dimensional $O(N)/O(N-1)$ nonlinear sigma model at large $N$, where it can be solved exactly. 
One can choose to preserve either dilatations together with rotations or translations in 
two directions, but not all of these symmetries. If one forces the preservation 
of all these symmetries in the quantum theory, it would end up being free. 
The partition function of the model is 
\begin{equation}
Z=\int [d{\vec \phi} ] [d\sigma] e^{-S} ~, ~
S =\frac{1}{2} \int d^2x \left [(\partial_\alpha {\vec \phi})^2+i\sigma [ {\vec \phi}^2 - N/\lambda ] \right ] ~,
\end{equation}
where  $\vec \phi = (\phi_1, \phi_2 ,\cdots, \phi_N)$ is the N-component vector field and the
limit $N\to\infty$ is taken at fixed $\lambda$, with $\lambda \equiv g^2 N$. The auxiliary
field $\sigma$ enforces the constraint ${\vec \phi}^2 \equiv \vec \phi \cdot \vec \phi =1/g^2$. 

\subsection{Standard, translation invariant analysis}

At large $N$, $Z$ can be computed by saddle point. 
An implicit cutoff is assumed -- it will be made explicit later on. 
In the standard approach, the global
$O(N)$ symmetry, which is nonlinearly realized at the classical level, becomes linearly realized at the quantum level
by a multiplet of $N$ massive scalar particles. Their mass, $\mu$, is
obtained by seeking a translation invariant saddle point of the form
$\sigma(x) = -i\mu^2$.
In Fourier space we have at the saddle: 
\begin{equation}
{\vec \phi}(x)=\frac{1}{(2\pi)^2} \int d^2p e^{ipx} {\hat{\vec\phi}}(p) ~,~
S = \frac{1}{2}\int \frac{d^2p}{(2\pi)^2} \hat{\vec \phi}^*(p) (p^2+\mu^2) \hat{\vec\phi}(p) -\frac{1}{2\lambda}\mu^2 N
\end{equation}
with $\hat{\vec \phi}^*(p) \equiv \hat{\vec\phi}(-p)$. 
After integrating over the scalar fields we obtain $Z=\int d \mu^2 e^{-N F(\mu^2,\lambda)}$.
The value of $\mu^2$ is obtained by minimizing $F(\mu^2,\lambda)= \frac{1}{2}\int \frac{d^2p}{(2\pi)^2} 
\ln (p^2+\mu^2) -\frac{1}{2\lambda}\mu^2$
and is 
$\mu = \frac{\Lambda}{\sqrt{e^{4\pi/\lambda}-1}}$
for a UV cutoff of the form $p^2 < \Lambda^2$, where $\Lambda$
is a large momentum cutoff of mass dimension. For large $\Lambda/\mu$ we have
$\lambda \sim \frac{2\pi}{\ln( \Lambda/\mu )}$. 
This is the standard result with dimensional $\Lambda$ and $\mu$.

\subsection{Dilatation invariant analysis}

Following~\cite{Fubini:1972mf}  we first go to radial coordinates, $x_1=r\cos\theta,~x_2 = r\sin\theta$
and then to Fourier space in $\theta$:
\begin{equation}
{\vec \varphi}(r,\theta) 
= \frac{1}{2\pi}~\sum_{m=-\infty}^\infty \hat{\vec\varphi}_m(r) e^{im\theta} ~,~
\hat{\vec\varphi}_{-m}(r) \equiv \hat{\vec\varphi}_m^*
\end{equation}
and 
\begin{equation}
\sigma(x) = \frac{1}{2\pi}\sum_{m=-\infty}^\infty
\frac{\hat\sigma_m(r)}{r^2} e^{im\theta} ~,~ \hat\sigma_{-m}(r) \equiv \hat\sigma^*_m(r) ~ .
\end{equation}
We now map the radial coordinate to the infinite line. This step
turns dilatations into translations in $t$ and inversion
into a parity in $t$. 
\begin{equation}
r=e^t, ~~ dr=e^tdt, ~~{\vec \Phi}_m(t) \equiv \hat{\vec\varphi}_m(r), ~~
\Sigma_m(t) \equiv \hat\sigma_m(r)
\end{equation}
In the new variables the classical action is
\begin{equation}
\frac{1}{4\pi}
\int_{-\infty}^\infty dt dt'
\sum_{mm'}{\vec \Phi}_m(t) D^{mm'}(t,t') {\vec\Phi}_{m'}(t')
- \frac{i}{2\lambda} N \int_{-\infty}^\infty dt \Sigma_0(t)
\end{equation}
with
\begin{equation}
D^{mm'}(t,t') = 
\left[ \left( -\frac{d^2}{dt^2} + m^2\right) \delta_{m+m',0}
+\frac{i}{2\pi}\Sigma_{-m-m'}(t)\right] \delta(t-t') ~ .
\end{equation}
We now look for a rotation invariant saddle point that also is translation invariant in $t$, 
$\Sigma_m(t) = 2\pi \bar\Sigma \delta_{m,0}$. 
In Cartesian coordinates, such a saddle is scale invariant, but singles out one point, serving as the origin. 

After integrating out the vector field, we obtain the following unregulated saddle point equation: 
\begin{equation}
\int_{-\infty}^\infty dk \sum_{m=-\infty}^\infty 
\frac{1}{k^2+m^2+i\bar\Sigma} =  \frac{4\pi^2}{\lambda}
\end{equation}
The sum over $m$ is unrestricted and can be done in closed form, producing 
\begin{equation}
\frac{1}{4\pi}\int_{-\infty}^\infty dk \frac{\coth \pi\sqrt{k^2+i\bar\Sigma}}
{\sqrt{k^2+i\bar\Sigma}} = \frac{1}{\lambda}
\end{equation}
We now set $i\bar\Sigma = \mu^2$ and choose a cutoff $\Lambda > 0$. 
$\Lambda > 0$ is dimensionless since $t$ was dimensionless and $k$ is the
conjugate momentum to $t$. We regulate by integrating over $-\Lambda < k < \Lambda$ assuming $\Lambda/\mu, \Lambda/\mu_0 \gg 1$. 
This preserves also the $t\to -t$ symmetry. 
\begin{equation}
\frac{1}{4\pi} \int_{-\Lambda}^\Lambda dk 
\left[ \frac{\coth \pi\sqrt{k^2+\mu^2}}{\sqrt{k^2+\mu^2}} -\frac{1}{\sqrt{k^2+\mu_0^2}}\right] 
=   \frac{1}{\lambda} - 
\frac{1}{2\pi} \sinh^{-1}\frac{\Lambda}{\mu_0} \equiv  \frac{1}{\lambda_R(\mu,\mu_0)} \ge 0
\label{sp}
\end{equation}
$\lambda_R$ is a renormalized coupling. $\mu$ and $\mu_0$ are dimensionless.
$\mu_0$ is an arbitrary renormalization point. The bare coupling depends on 
the cutoff in the standard way, except the cutoff is dimensionless: 
$\lambda \sim \frac{2\pi}{\ln ( 2\Lambda/\mu_0 )}$.
A nonzero saddle point will always be found. 
The saddle point value $\mu$ never exceeds a maximal value, which depends on $\mu_0$ and is 
defined by
\begin{equation}
 \mu_{\rm max}(\mu_0):\ \ \ 
\int_{-\infty}^\infty dk 
\left[ \frac{\coth \pi\sqrt{k^2+\mu_{\rm max}^2(\mu_0)}}{\sqrt{k^2+\mu_{\rm max}^2(\mu_0)}} -\frac{1}{\sqrt{k^2+\mu_0^2}}\right]
=  0  ~ .
\end{equation}

The propagator of the scalar field is
\begin{equation}
\langle {\Phi^*}_{m'}^i(t) \Phi^j_m(t')\rangle =
\delta^{ij} \delta_{mm'}
\int_{-\infty}^\infty \frac{dk}{2\pi}
\frac{e^{ik(t-t')}}{k^2+m^2+\mu^2} 
=\delta_{ij} \delta^{mm'}
\frac{e^{-\sqrt{m^2+\mu^2}|t-t'|}}{2\sqrt{m^2+\mu^2}} ~ .
\label{corr1}
\end{equation}

\subsection{Cartesian Translation Invariance}

Reading off 
the set of dilatation eigenvalues displayed by eq.~(\ref{corr1})  
we see that it does not contain an equally spaced ladder. 
Consequently, we cannot construct translation generators satisfying the correct
commutation relations~\cite{frad} with dilatations in the sector generated by the action 
of $\vec \Phi$ on the vacuum. 
The deviation of the dilatation spectrum from equal spacing
is small if $m \gg \mu$. There is  
approximate translation invariance for angular separations that are small relatively to $\frac{1}{\mu}$. 
Such separations can be resolved only by waves with high enough angular momenta $m$. 
Because inversion has also been
preserved in the quantization, if translations could be realized, special conformal transformations
would come in automatically and the full conformal group would be realized. 
Because of rotation invariance, only
one linear combination of translations needs to be considered in detail. 

Comparing the manipulations we did here to those we would have done had we declared we are 
interested in the traditional view of the model at finite non-zero temperature, we realize that
while the language has changed, the equations have not. There exists only one bosonic vector
$O(N)/O(N-1)$ sigma model on $\mathbb S\times \mathbb R$. In traditional 
terms, the circumference of $\mathbb S$ in units of the dynamically generated mass at zero temperature is 
a free parameter. Insisting that $\mathbb S$ is a geometric circle made its length equal to $2\pi$ in
the untraditional view. 
In traditional terms the propagator decays exponentially in the Cartesian 
spacetime directions, while in untraditional terms the propagator decays exponentially 
in the coordinate conjugate to the dilatation generator.  

\subsection{One useful lesson}

We did not get anything truly new in the untraditional quantization; the ``novelty'' boiled
down to semantics. 
But, we learned a lesson which will be useful in the following: 
if we started from a classical conformal theory that allowed this symmetry 
to be fully preserved at the quantum level and chose to preserve dilatations and rotations 
explicitly by the cutoff
procedure, translations would be restored in the continuum limit. 
The generators of translations could be reconstructed from the commutation relations of the conformal algebra
using the available generators of rotations, dilatation and inversion~\cite{frad}. Eventually we would
get the same theory we would have gotten from a traditional lattice regularization that preserved 
a discrete version of translations from the start 
but left dilatations and rotations unpreserved. However, at the lattice regulated level, the choice to preserve 
dilatations seems more efficient.

\section{Lattice radial quantization in three dimensions}

In the continuum we would be dealing with a theory on $\mathbb{S}^2\times\mathbb{R}$. 
The critical exponents of the same CFT on $\mathbb{R}^3$ can be extracted from the
dilatation operator on $\mathbb{S}^2\times\mathbb{R}$~\cite{Cardy:1985}. In the
two dimensional case one has $\mathbb{S}\times\mathbb{R}$ and the same can be done~\cite{Cardy:1984rp}.
This works well because the factor $\mathbb{S}$ is easy to discretize. In three 
dimensions we need to discretize $\mathbb{S}^2$ and it becomes impossible to replace the continuum 
$O(3)$ symmetry by symmetries under ever larger discrete subgroups as the lattice is refined. 
One is limited to a few discrete subgroups. Hopefully, this gets compensated by their stronger, 
nonabelian structure. One can latticize $\mathbb{S}^2$~\cite{Cardy:1985} 
by the set of vertices of a platonic
solid, but
this does not work well~\cite{Alcaraz:1987}. A way to improve on this is to use 
regular lattice refinements of the faces of the platonic solid~\cite{Weigel:2000ee}. 
For a cube, this approach produced reasonably accurate exponents in the $Z_2$ even sector of the Ising model.
However, the estimate of the exponent $\eta$ in the odd sector had a substantially larger deviation from the known 
value~\cite{Pelissetto:2000ek}.  
 Our numerical experiment will exploit the lesson we learned in the previous section 
on a lattice similar to that of~\cite{Weigel:2000ee}. 

To have as much symmetry as possible we employ a closed two dimensional surface $\Sigma$,  
consisting of the boundary of an icosahedral solid instead of a cube.   
The linear distances between points on $\Sigma$ and the origin
are bounded to a relatively narrow interval on the positive axis. 
The icosahedron has 12 vertices and 20 faces given by identical flat
equilateral triangles.  Its symmetry group $I_h$ is a 120 element subgroup of
$O(3)$. The angular momenta $l=0,1,2$ representations of $O(3)$ remain
irreducible representations under $I_h$.  This is enough to isolate
a scalar primary state and two of its immediate descendant states on the basis of their 
behavior under $I_h$. $\mathbb{R}^3$ is visualized as a set of 20 infinite triangular pyramids 
glued together with the apexes meeting at the origin. In isolation, each pyramid can be 
(non-conformally) mapped into a uniform prism with an equilateral triangle as cross section. 
These prisms are then glued together in a manner inherited from the pyramids. We end up 
with $\Sigma\times\mathbb{R}$.

We regularize $\Sigma$ by replacing each face by a triangular piece of a
regular two dimensional triangular lattice. Consequently, each side of the
original triangle has extra $s-1$ equally spaced points placed on it ($s \ge
1$).  The number of sites on one icosahedral shell is $ 2 + 10 s^2 $. 
Identical grids replacing $\Sigma$ are strung along an equally spaced
infinite straight line, making up a cylinder. Matching vertices on neighboring
$\Sigma$-slices are connected by links of equal lengths.

For a finite cylinder, the partition function is of the usual form, 
\begin{equation}
Z = Tr_\sigma  e^{\textstyle  \sum_{t,i}  \beta \sigma(t,i) \sigma(t+1,i)+\sum_{t,\langle i j\rangle} \beta  \sigma(t,i)\sigma(t,j)}  \; ,
\end{equation}
where $\langle i j\rangle$ denotes a nearest-neighbor pair on the same 
icosahedral shell and $t = 0, \cdots T-1$ labels the shells along the cylinder.
The trace is a sum over the Ising spin values $\sigma(t,i) = \pm 1$ for all sites $(t,i)$.


To get to the continuum limit of the Wilson-Fisher conformal field theory, we
need to tune $\beta$ to its infinite $s$ critical point.  The coupling $\beta$
has been chosen the same in all directions for simplicity. 
The intra-shell lattice spacing and the lattice spacing on the
triangulated shells are fixed by demanding that descendants have equal unit spacing
as we approach the continuum limit. This is a self-consistent definition only if the descendant 
``masses'' fall on a straight line. This happens only at 
the critical value of $\beta$. 
To get information on
the spectrum of the transfer matrix it is convenient to compactify the
infinite axis of the cylinder to a circle with periodic boundary conditions on
the spins. We chose our cylinders to have lengths $T$ which scale with the refinement as 
$T = \rho  s$. 

Our focus is on the 3D Ising critical exponent $\eta$. This number is
determined by the dimension $\Delta_\sigma$ of the lowest primary in the $Z_2$ odd
sector of the space the transfer matrix acts on.  
To get the primary state and some of its descendants we define field variables
on each $t$-labeled shell:
\begin{equation}
{\hat \sigma} (t, l,m) = \sum_i \Delta\Omega_i  Y_l^m (\theta_i,\phi_i)
\sigma(t,\theta_i,\phi_i) \; .
\end{equation}
$\Sigma$ is parameterized by the polar and azimuthal angles $\theta,\phi$. 
$t$ corresponds to the logarithm of the amount 
of dilatation a specific $\Sigma$ embedded in $\mathbb{R}^3$ would have undergone.
The $Y_l^m (\theta,\phi)$ are 
the standard spherical harmonics with $l=0,1,2$. The measure $\Delta \Omega_i$ 
is computed from the area of the spherical triangles adjacent to $\theta_i,\phi_i$ 
projecting onto a unit sphere in order to assure a rapid convergence to 
the orthonormality condition for the $Y_l^m (\theta,\phi)$'s in the continuum.
The $I_h$ symmetries and dilatations 
guarantee $\langle {\hat\sigma}(t, l,m){\hat\sigma}(t',l',m')\rangle=C_l (t-t') 
\delta_{ll'}\delta_{mm'}$ for $l,l'=0,1,2$ with  $|m|\le l$ , $
|m'|\le l'$.  From the $l=0$ channel we get the
dimension of the primary state and from the $l=1,2$-channels the dimensions of
the next two descendant states.  We denote these three eigenvalues
of the dilatation operator by $\Delta_l$, where $\Delta_l$ plays the role of a
mass when $t$ is regarded as a time variable. 
The eigenvalues of the lattice dilatation operator corresponding to the
continuum dimensionless numbers $\Delta_l$ are denoted by $\mu_l$.  The theory
now has an ultraviolet cutoff denoted by $\cut$. $\cut^{-1}$ is dimensionless,
determining the smallest angular separation on the discretized $\Sigma$. 
\begin{equation}
\Delta_l=\mu_l \cut
\end{equation}
$\cut\propto s$ as $s\to\infty$ and we anticipate an 
integer spacing of descendants,  $\Delta_l=\Delta_0+l$, in the $s\to\infty$ 
limit.  Having three values of $\mu_l$ at our disposal due to the icosahedral
symmetry, we can test numerically whether they indeed fall on one straight line.  The
slope of this line determines $\cut$ and fixes the ratio of the lattice
spacing on the sphere relative to radial axis in  $t$.  The $\mu_l$ will go as
$s^{-1}$ at large $s$ and the standard critical exponent $\eta$ is given by
the anomalous portion of $\Delta_l$ for any $l$: $\eta=2(\Delta_l-l-1/2)$.
 
It is known that $\eta$ is small relative to 1~\cite{Pelissetto:2000ek}. This puts a high accuracy
demand on the $\mu_l$'s, which are the directly measurable quantities. 

\section{Results}

\begin{figure}[ht]
\centering
\hfil
\includegraphics[width=0.45\textwidth]{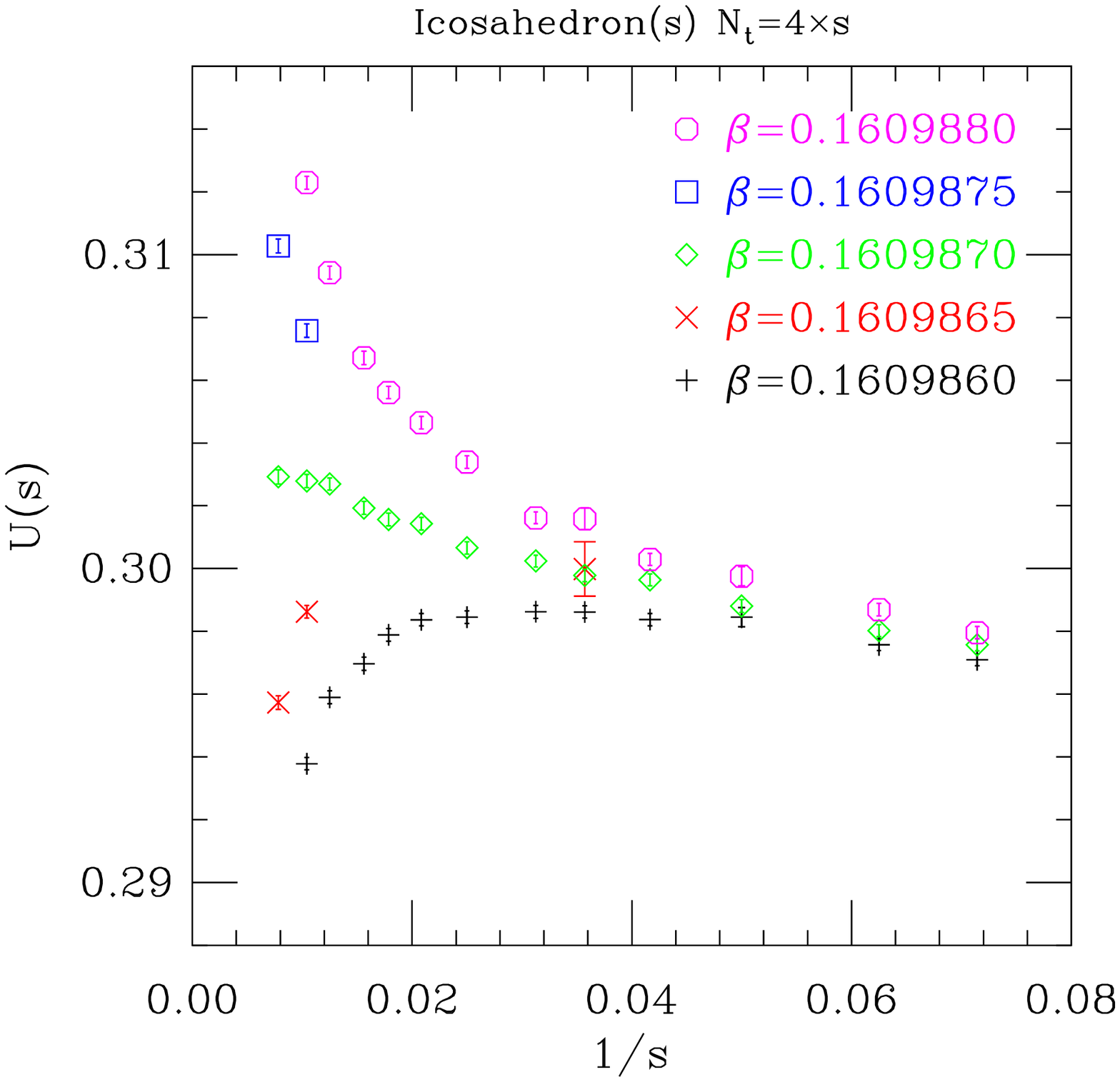}
\hfil
\includegraphics[width=0.45\textwidth]{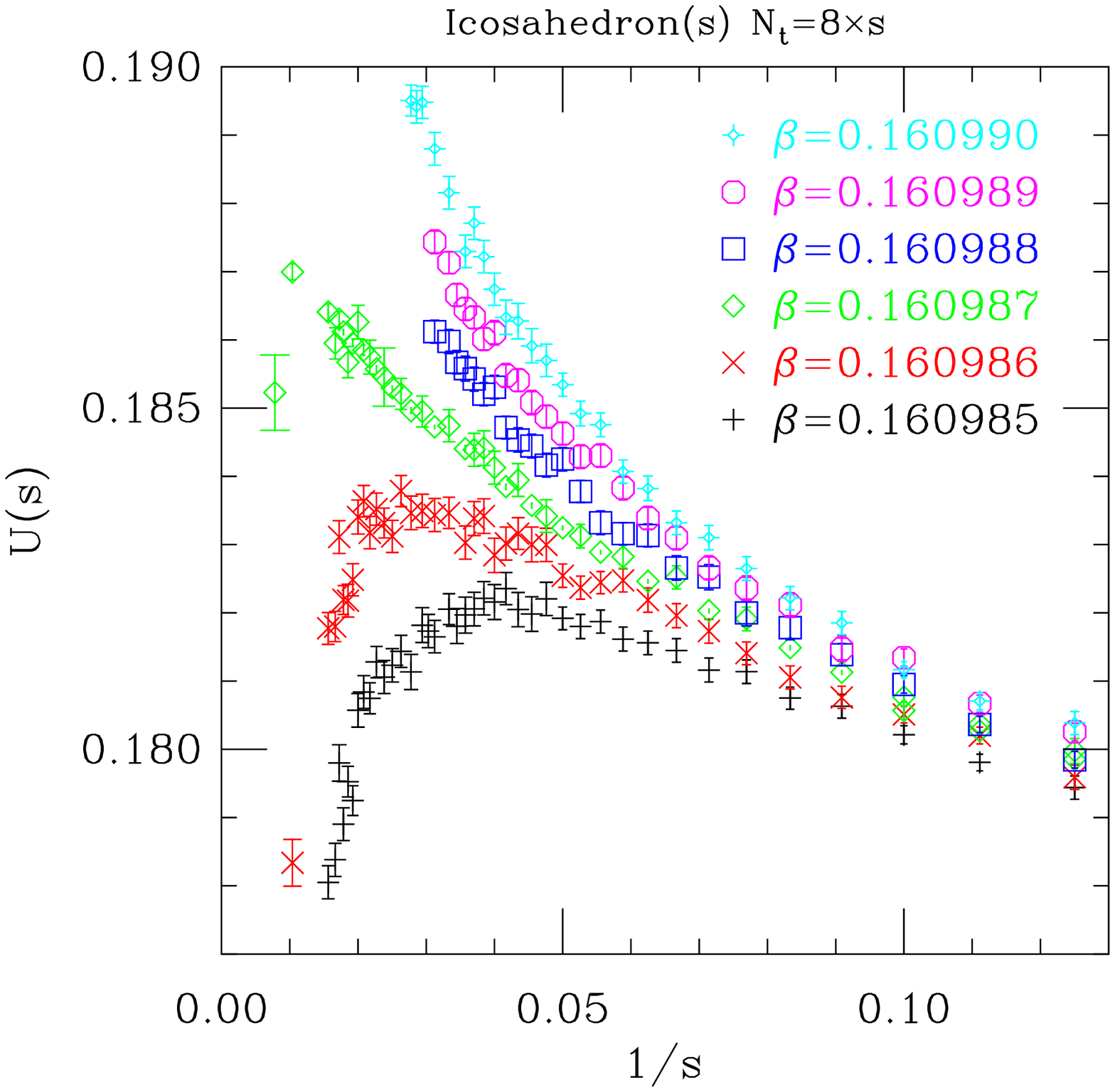}
\hfil
\caption{\label{fig:Binder}Determining $\beta_c$ from the Binder cumulants
$U(s) = 1 - \langle M^4\rangle /3 \langle M^2 \rangle^2$ near the
pseudo-critical point for two different aspect ratios $\rho$ and increasing
values of $s$.}
\end{figure}

First the critical point $\beta_c$ was determined  by
computing the 4th-order Binder cumulant~\cite{Binder:1981sa}, 
\begin{equation}
\label{eq:Binder_cumulant}
U(\beta,s,\rho) = 1 - \frac{\langle M^4 \rangle}{3 \langle M^2 \rangle^2} \; .
\end{equation}
Our initial computations of moments of the magnetization were performed using
the Swendsen-Wang cluster algorithm~\cite{Swendsen:1987ce}.  Later, we
switched to the more efficient~\cite{Tamayo:1990} single cluster Wolff
algorithm~\cite{Wolff:1988uh}.  To locate the critical point, we used aspect
ratios $\rho = 4, 8$.  with  $s \ge 16$ and $\beta$ such that $|\beta - \beta_c|
s^{1/\nu} \le 0.0012$.  We then fit the data to:
\begin{equation}
U(\beta,s,\rho) \simeq U(\beta_c,\infty, \rho) + a_1(\rho) [\beta - \beta_c]
s^{1/\nu} + b_1(\rho) s^{-\omega} .
\end{equation}
Subsets of the data used appear in Fig.~\ref{fig:Binder}.  The best fit gave
$\beta_c$ = 0.16098691(1) with critical Binder cumulants $U(\beta_c,\infty,4)$
= 0.3032(1) and $U(\beta_c,\infty,8)$ = 0.18765(5) with a $\chi^2/\mathrm{dof}
= 1.37$, $\mathrm{dof} = 212$.  The best-fit exponents $\nu = 0.621(4)$ and
$\omega = 0.797(6)$ are consistent with the 
values in~\cite{Pelissetto:2000ek} at the 2-sigma level.

To compute the  correlation functions, 
$C_l(t) = \langle {\hat\sigma}(t + t_0, l,m){\hat\sigma}(t_0,l,m)\rangle$ , 
we generated ensembles at
$\beta = 0.160987$.  Our final results for $\beta_c$ from the Binder cumulant
suggest this might be slightly in the ordered phase as $s\to\infty$.  Each
independent run was thermalized using 2048 sweeps of the Wolff algorithm
followed by 8192 sweeps with one estimate of the 
correlation function after each sweep.  We defined a sweep to be $19 s / 2$
Wolff cluster updates.  This choice sets the average number of spins flipped
per sweep to be about equal to the total volume.  All results for a given run
are then averaged together to form a single blocked estimate and many
thousands of independent block estimates are combined to form the ensemble at
each $s$, as shown in Table~\ref{tab:runs}.  The jackknife method was used to
estimate errors.

\begin{table}[ht]
\centering
\begin{tabular}{|c|c|c|c|c|c|c|c|c|c|c|c|}
\hline
$s$ & 8     & 9     & 10    & 11    & 12    & 14    & 16    & 18    & 20    & 22    & 24    \\
\hline
$N$ & 19456 & 19999 & 20480 & 20000 & 18975 & 20480 & 23552 & 22528 & 19328 & 10624 & 10202 \\
\hline\hline
$s$ & 28    & 32    & 36    & 40    & 44    & 48    & 52    & 56    & 58    & 64    &       \\
\hline
$N$ & 6656  & 25792 & 8832  & 15168 & 13608 & 2528  & 7935  & 8017  & 1088  & 1000  &       \\
\hline
\end{tabular}
\caption{\label{tab:runs} $N$ is the number of independent runs used for the
correlation function at each refinement $s$. Other details
of the runs are described in the text.}
\end{table}

\begin{figure}[ht]
\hfil
\includegraphics[width=0.45\textwidth]{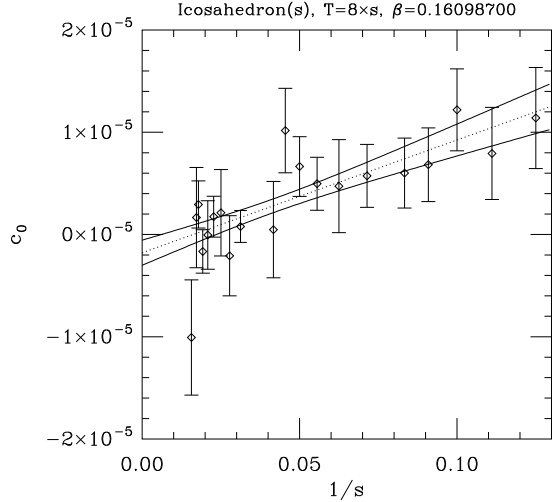}
\hfil
\caption{\label{fig:discb16098700_fit} The disconnected contribution to the
correlation function as described in the text.}
\end{figure}

We found it very useful to evaluate the correlation
function using the momentum-space single cluster improved estimator
method~\cite{Ruge:1993, Ruge:1994jc}.  As we are only interested in the
rotation-invariant part of the correlation function on any given lattice,
we also average over the $m=-l,...,l$ states in the various representations
labeled by $l$.  Since our value of $\beta$ should be slightly larger than the
pseudo-critical coupling at any finite $s$, we expect that our $l = 0$
correlation function will have a small disconnected contribution, which should
vanish as $s \to \infty$, assuming $\beta = \beta_c$.  If the disconnected
contribution remained non-zero, it  would suggest that $\beta > \beta_c$; 
otherwise, we would conclude that $\beta < \beta_c$.
The contribution of the lowest eigenstate of the transfer matrix to the
connected correlation function in momentum space is given by 
\begin{equation}
\widetilde{C}_l(k) = c_0 \delta_{l,0} \delta_{k,0} + a_l
\frac{\left(1-e^{-\mu_l T}\right) \sinh(\mu_l)}{\sinh^2(\mu_l/2) +
\sin^2(k/2)} \; ,
\end{equation}
where $k=2 \pi n / T$ with $n = 0, \cdots T-1$ is the momentum conjugate to $t$ along
the cylinder and we have included the disconnected contribution $c_0$.  We can
determine $c_0$ by fitting $\widetilde{C}_0(k)$ for $k \ne 0$ and subtracting
a smooth extrapolation of $\lim_{k \to 0} \widetilde{C}_0(k)$.  We found that
our data required parameterizing the lowest eigenvalue plus at least two higher
eigenvalues. We got excellent fits with $\chi^2 / \mbox{dof} \lesssim 1$ and
estimates of the lowest eigenvalues which are essentially free of
contamination.  The result is shown in Fig.~\ref{fig:discb16098700_fit}.  A
linear extrapolation in $1/s$ indicates that the disconnected contribution
vanishes at finite $s \approx 64$. Then, $\beta = 0.160987 < \beta_c$, in
apparent contradiction to the Binder cumulant result.

\begin{figure}[ht]
\centering
\hfil
\includegraphics[width=0.45\textwidth]{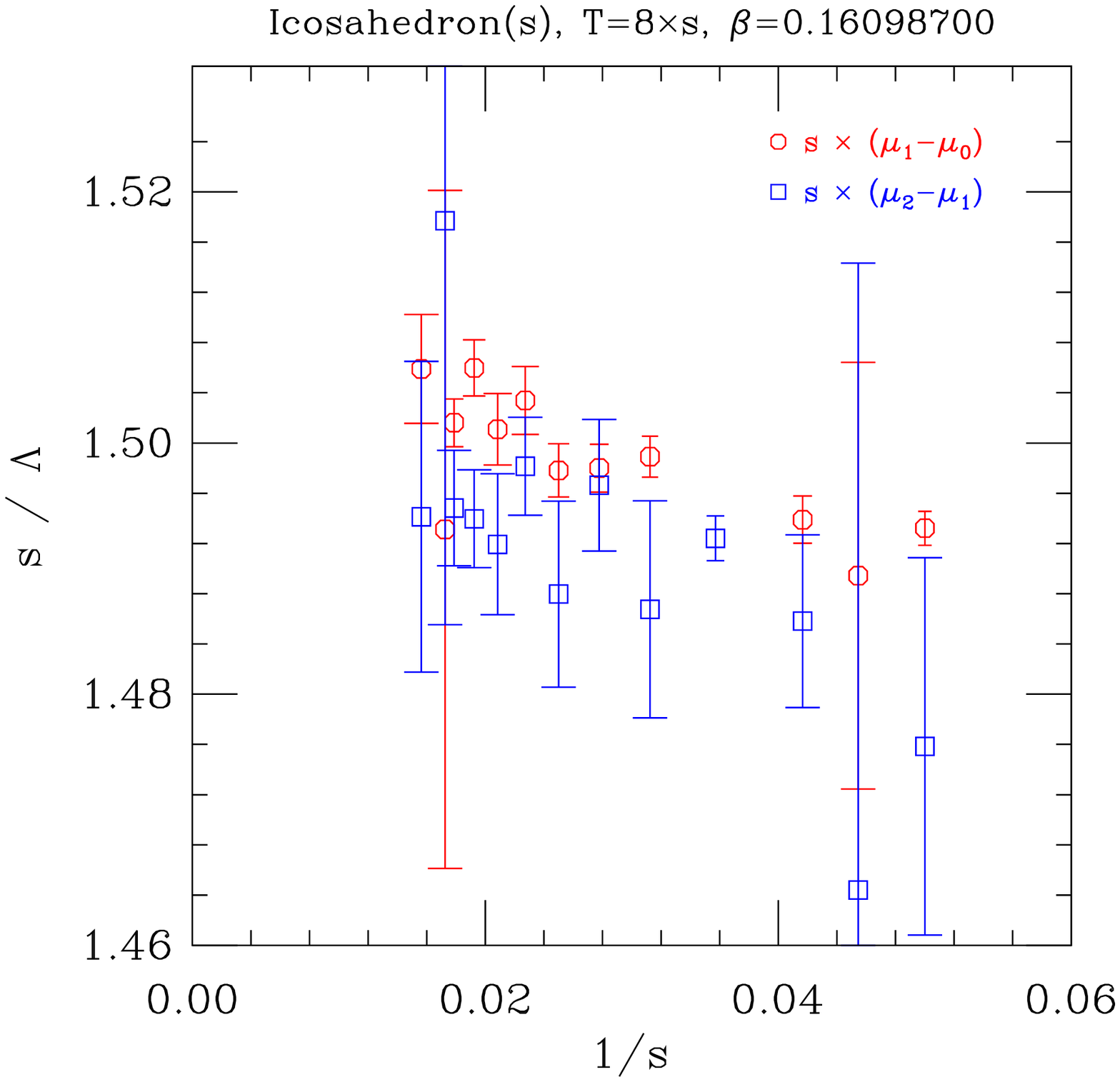}
\hfil
\includegraphics[width=0.45\textwidth]{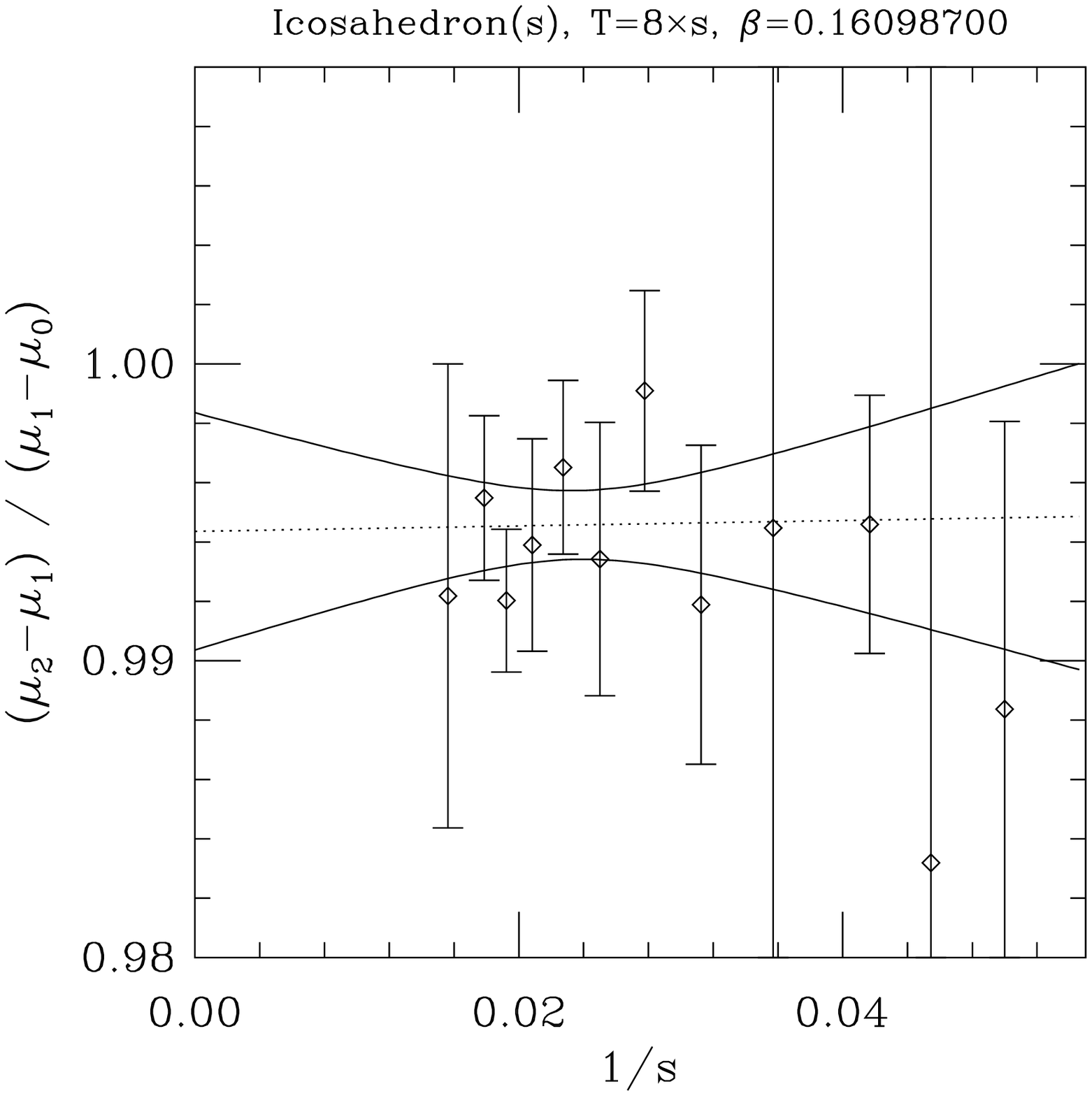}
\hfil
\caption{\label{fig:scaling_spacing} The left figure shows the scaling of
$\Lambda$ with $s$.  The extrapolated value is 1.51(1) with the
uncertainty dominated by the systematic difference between the two estimates.
The right figure checks whether the $\mu_l$ fall on a straight line.  
We fit to a linear function and find an intercept of 0.994(4) and
slope of 0.0(2) with $\chi^2/\mbox{dof} = 0.43$ for 11 dof.}
\end{figure}

\begin{figure}[ht]
\centering
\includegraphics[width=0.45\textwidth]{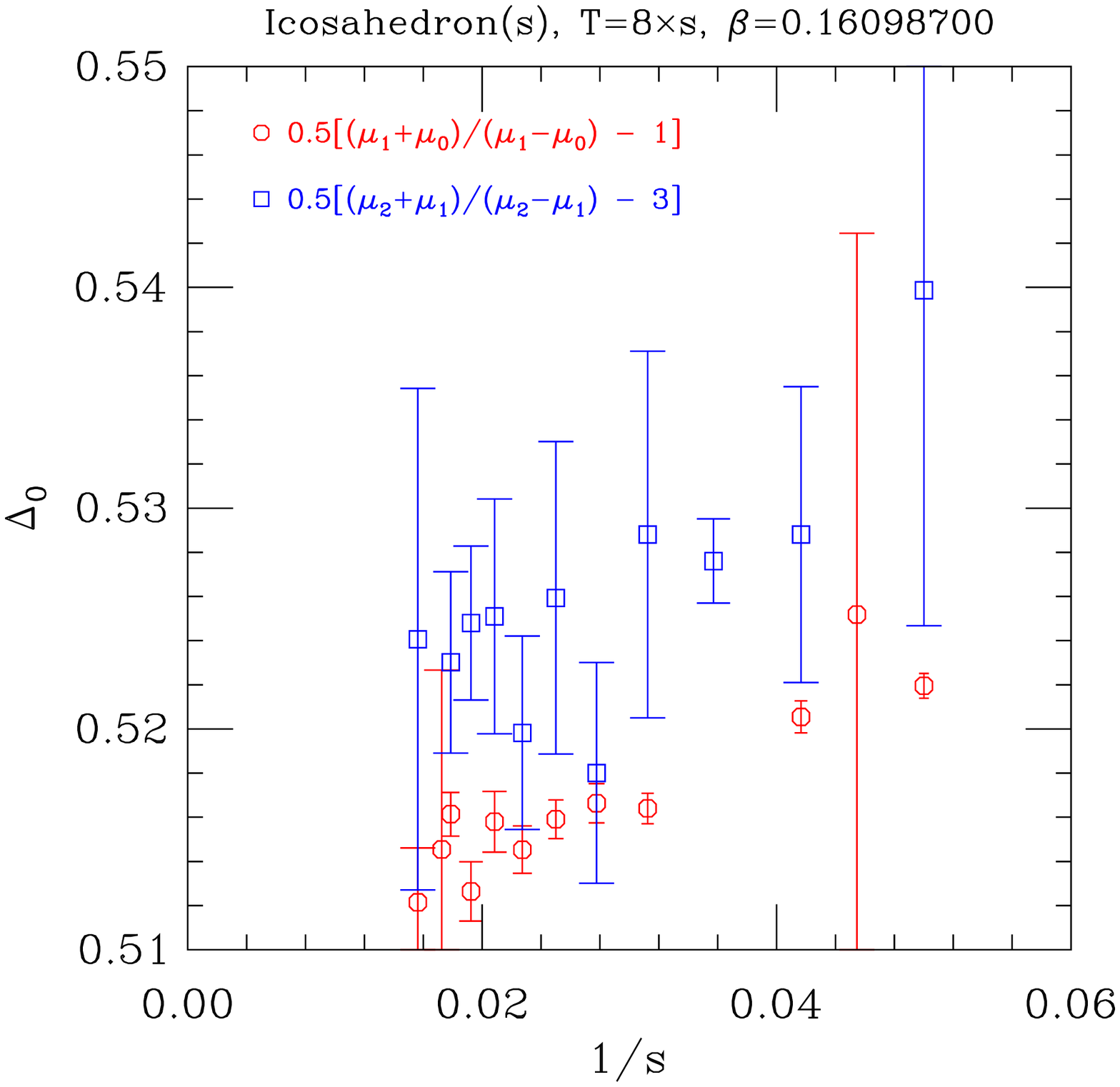}
\caption{\label{fig:delta} The scaling exponent of lowest $Z_2$-odd primary
operator \textit{vs.}\ $1/s$.  The lower (red) points come from $\mu_0$ and $\mu_1$.
The determination of $\mu_0$ is unreliable because of the
disconnected contribution to the correlator,  given the uncertainty whether $\beta_c < 0.160987$ 
$\beta_c > 0.160987$. A linear extrapolation from upper (blue) points, based on $\mu_1$ and $\mu_2$, gives $\Delta_0 =
0.517(5)$, which is consistent with the estimate
0.5182(3) in~\cite{Pelissetto:2000ek}.}
\end{figure}

We relate the $\mu_l$'s for $l=0,1,2$ to the
eigenvalues of the dilatation operator by 
$\mu_l = \Lambda^{-1} [ \Delta_0 + l ]$ where $\Lambda^{-1} \simeq c_1 / s $
as $s \to \infty$.  We find $c_1 \approx 1.51(1)$ with the
uncertainty dominated by systematic error.  The the left figure of
Fig.~\ref{fig:scaling_spacing} shows evidence for also 
sub-leading contributions, $\mathcal{O}(1/s^2)$.  We confirm the equal spacing rule of
descendants  by examining the ratios, $(\mu_{l+2} - \mu_{l+1}) / (\mu_{l+1} -
\mu_l)$, as shown on the right in Fig.~\ref{fig:scaling_spacing}.  We are 
now justified to estimate the scaling dimension of the
primary operator using 
\begin{equation}
\Delta_0 = \frac{l-l^\prime}{2} \left[
  \frac{\mu_{l} + \mu_{l^\prime}}{\mu_{l} - \mu_{l^\prime}}
  - \frac{l + l^\prime}{l - l^\prime}
\right]
\end{equation}
as shown in Fig.~\ref{fig:delta}. The determination is independent of $\Lambda$. 

A quite conservative estimate is 
$\eta=0.034(10)$, consistent with other estimates \cite{Pelissetto:2000ek}.
Our numerical study  could be extended to include additional primary operators
in both the $Z_2$-odd and $Z_2$-even sectors as well as a direct test of the
restoration of full conformal symmetry for 2- and 3-point correlators. 
\begin{figure}[ht]
\centering
\includegraphics[width=0.45\textwidth]{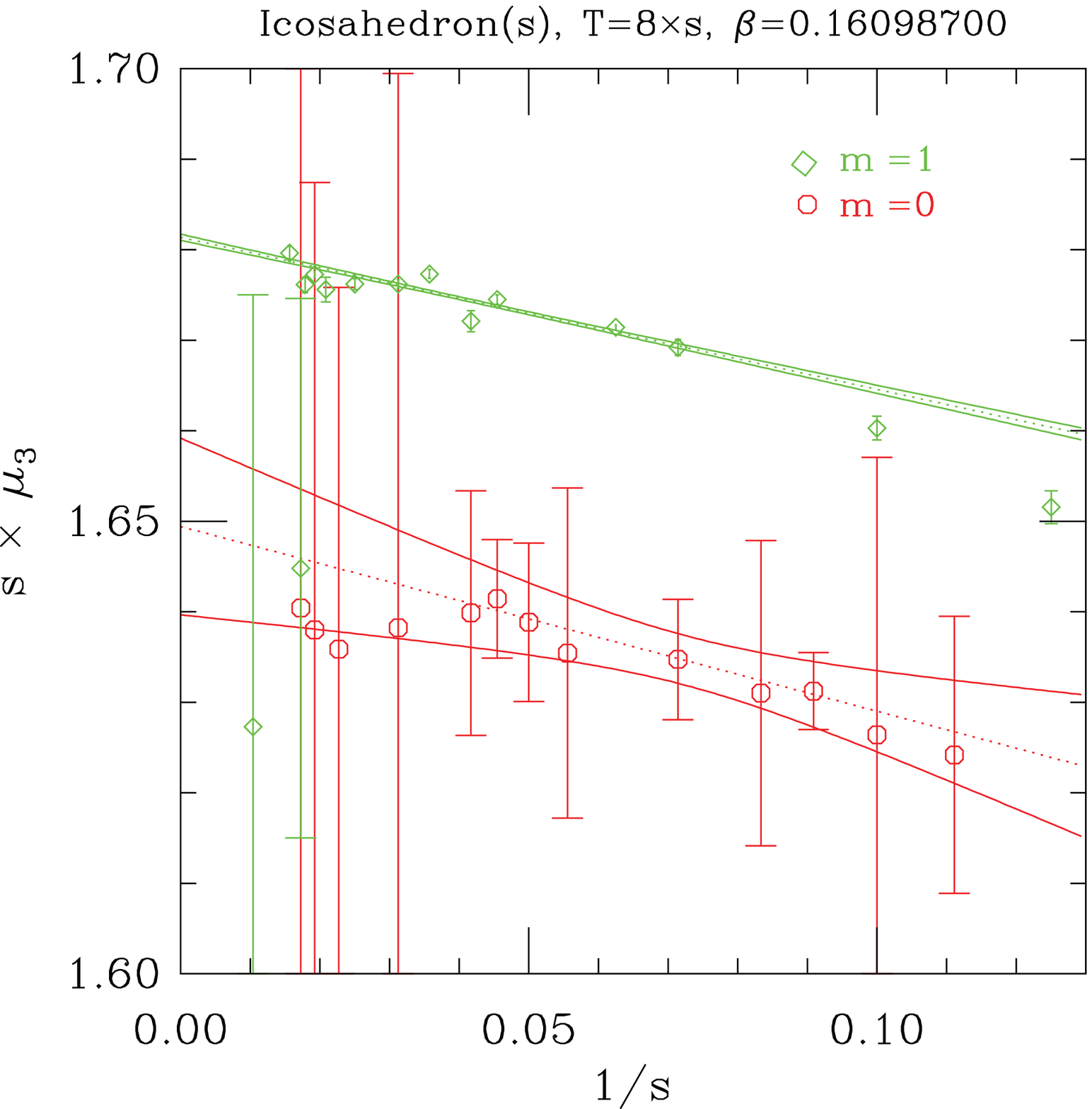}
\caption{\label{fig:mu1_b16098700_l3_v1} The eigenvalues $\mu_3$ for the $l = 3,
m = 0, \pm 1$ descendents of the lowest $Z_2$-odd 
operator \textit{vs.}\ $1/s$.  
The lower (red) circles are $m = 0$ and the upper (green) diamonds are $m = \pm 1$.  
The difference is a measure of the
breaking of $O(3$) symmetry at that $s$. }
\end{figure}
However as  a cautionary note, we  now briefly describe the behavior of 
the eigenvalues for the descendants with $l = 3$.

The 7-dimensional $l=3$ irrep of O(3) splits under $I_h$ into the sum of the 4-dimensional irrep
$G$ and the 3-dimensional irrep $T_2$. The $l=3$ states 
$m = \pm 2, \pm 3$ appear in both irreps of $I_h$ but the states $m = \pm 1$
only appear in $G$, while the state $m = 0$ only appears in $T_2$.  In
Fig.~\ref{fig:mu1_b16098700_l3_v1} we show the $l=3$ eigenvalues labeled by $m
= 0, \pm 1$.  If they extrapolated to the same value as $s \to \infty$, it
would be an  indication that O(3) symmetry is restored in the continuum
limit. The numerical evidence so far suggests the contrary, indicating that the continuum limit
with the icosahedral shell is not fully reproducing the conformal fixed 
Wilson-Fisher fixed point of the 3D Ising model on flat three-space. More refined methods are needed, 
as suggested below.

\section{Discussion and Outlook}

The method of lattice radial quantization presented in this paper holds
promise as a practical nonperturbative tool for conformal field theory.
We plan additional numerical and theoretical studies to realize this goal.

An  important theoretical question remains: How do exponents associated with
$l=0,1,2$ in our icosahedral shell system at $s\to\infty$ relate in principle to their exact $\mathbb{R}^3$ values 
in the 3D Ising model?  The numbers
in~\cite{Weigel:2000ee} and our $\eta$-value are consistent with equality, 
but  are far from providing overwhelming numerical evidence. Indeed, 
the apparent lack of convergence to a single $O(3)$  irreducible representation for $l=3$
in Fig.~\ref{fig:mu1_b16098700_l3_v1} suggests that the continuum limit
of our radial-icosahedral model is not equivalent to the 3D Ising  model
at the Wilson-Fisher fixed point on flat three space. A failure to reproduce the correct 
spectral degeneracy of $l > 2$ 
descendants at $s\to\infty$ would indicate distorted spectra for the primaries as well.
The icosahedral shell may be sufficiently spherical to keep these distortions in $\eta$  to
a level below our current estimate. 

We
are now studying in the continuum the consequence of quantization on  $\Sigma\times \mathbb{R}$ where
$\Sigma$ is sectionally flat. Also, 
we are implementing improved  actions for our lattice representation with nearest neighbor bonds
weighted to approach the continuum metric on  $\mathbb S^2 \times \mathbb R $,
similar to the discretization employed for our single spin operators.  
Both theoretical considerations and numerical tests are needed to clarify how a radial lattice 
can reproduce a conformal Euclidean field theory on flat space in the continuum limit. 

Lattice radial quantization should apply also to four dimensional gauge theories
with an amount of matter where the IR behavior is suspected to be controlled
by an interacting conformal theory. Historically, interest in such theories was motivated
by a search for realizations of walking Technicolor
scenarios~\cite{Appelquist:1986an} in their vicinity. Independently of the
relevance of the latter to Nature, there is theoretical interest in these
theories, in particular in the large $N$ limit for the gauge group $SU(N)$
because of the potential existence of a specific string dual in the IR. In the
case of 4D gauge theories, the anomalous dimensions of interest are typically
large, so the demands on numerical accuracy are weaker than those for $\eta$. 

\paragraph{Acknowledgments:}
RCB acknowledges support under DOE grants DE-FG02-91ER40676,
DE-FC02-06ER41440 and NSF grants OCI-0749317, OCI-0749202.  RCB has benefited
from conversations with Joseph Minaham.  GTF acknowledges partial support by
the NSF under grant NSF PHY-1100905.  RCB and GTF also  thank the Galileo
Galilei Institute for Theoretical Physics for the hospitality and INFN for
partial support during the workshop ``New Frontiers in Lattice Gauge
Theories''.  HN acknowledges partial support by the DOE under grant number
DE-FG02-01ER41165. HN is grateful for support under the Weston visiting
scientist program at the Weizmann Institute in the Faculty of Physics.  HN has
benefited from conversations with Micha Berkooz, Rajamani Narayanan and Adam
Schwimmer.

\end{document}